\documentclass[aps,prl,reprint,superscriptaddress,floatfix,nofootinbib,
  longbibliography]{revtex4-2}

\usepackage[T1]{fontenc}
\usepackage{lmodern}
\usepackage{amsmath,amssymb,bm}
\usepackage{graphicx}
\usepackage[svgnames]{xcolor}
\usepackage[colorlinks=true,citecolor=NavyBlue,linkcolor=NavyBlue,
  urlcolor=NavyBlue]{hyperref}
\graphicspath{{./}{figures/}}
\newcommand{\paperfigure}[2]{%
  \IfFileExists{#1}{\includegraphics[width=1.0\columnwidth]{#1}}{%
    \IfFileExists{figures/#1}{%
      \includegraphics[width=\columnwidth]{figures/#1}}{%
      \fbox{\parbox[c][#2][c]{0.94\columnwidth}{\centering
        \small Figure file not supplied\\[4pt]\texttt{#1}\\[4pt]
        Caption retained for manuscript review.}}}}}

\begin{document}

\title{Nonequilibrium Broken-Symmetry States from Retained Observables}

\author{Sankha Subhra Bakshi}
\affiliation{Department of Physics, University of Virginia,
Charlottesville, Virginia 22904, USA}
\author{Gia-Wei Chern}
\affiliation{Department of Physics, University of Virginia,
Charlottesville, Virginia 22904, USA}

\begin{abstract}
In equilibrium, broken-symmetry states are selected by the competition between
energy and entropy. Out of equilibrium, slowly relaxing observables retain
information about the preparation and may reshape this competition, allowing
orders that are not thermally favored. Identifying such states ordinarily
requires following slow, protocol-dependent dynamics. Here we recast this
dynamical selection as a static constrained variational problem. A small set
of experimentally measurable or numerically accessible quantities—particle
number, energy, and slow observables—defines a constraint space. The underlying dynamics determines where the system lies in this space, while entropy maximization over
both quasiparticle occupations and a prescribed class of candidate orders
determines which order is selected. Restricting the variational states to the
Gaussian manifold yields a computationally inexpensive constrained
Hartree--Fock implementation. For photoexcited Hubbard models, the approach
reproduces the long-time TDHF suppression of square-lattice
antiferromagnetism and identifies incommensurate spiral order in place of the
equilibrium $120^\circ$ state on the triangular lattice. For a continuously
driven ferromagnetic double-exchange model, evaluating the constrained order
map at the stationary NESS energy reveals a restricted region in which
N\'eel order is favored. The framework therefore turns the search for
nonequilibrium broken-symmetry states into a practical static screening
problem, without requiring simulations of every preparation protocol.
\end{abstract}


\maketitle

Equilibrium phase diagrams organize collective states in terms of a small set
of control parameters, such as temperature, pressure, and carrier density,
under the assumption that all microscopic degrees of freedom have relaxed to
a common thermal distribution. Ultrafast excitation provides a fundamentally
different route through the phase space of quantum materials: a short pulse
can redistribute quasiparticles, excite selected orbital or lattice sectors,
and transiently modify the interactions that stabilize an ordered phase
\cite{deLaTorre2021,Murakami2025}. Because these degrees of freedom relax on
widely separated timescales, the resulting state need not correspond to any
point on the equilibrium phase diagram. Time-resolved experiments have
revealed nonthermal melting, recovery, and persistence of charge order
\cite{Rohwer2011,Tomeljak2009,Lee2012,Beaud2014,Maklar2021}, together with
pump-induced orbital reorganization \cite{Tomimoto2003,Polli2007}. More
strikingly, optical excitation can create long-lived hidden states
\cite{Stojchevska2014,Ravnik2018,Vaskivskyi2016,Ravnik2021,Gao2022}, induce
or reconstruct charge order
\cite{Kogar2020,Cheng2022,Yoshikawa2021,Domrose2023}, enhance ferromagnetism
\cite{Disa2023}, and generate superconducting-like optical responses
\cite{Fausti2011,Mitrano2016,Budden2021}. These observations show that
excitation can do more than melt an equilibrium phase: \textit{it can redirect the
system toward a distinct collective order that is inaccessible by heating.}

After excitation, a system can relax without reaching thermal equilibrium.
Collisions can rapidly redistribute particles and energy, while band
populations, spin polarizations, or energy stored in weakly coupled sectors
change much more slowly. We assume that the fast degrees of freedom
equilibrate within the restrictions imposed by these slowly changing
quantities. This partial equilibration provides the physical basis for a
constrained maximum-entropy description. The evolution of order can
nevertheless be rich and strongly dependent on the preparation, involving
competing instabilities, nucleation, and domain growth. Even without
resolving this dynamics, a useful question remains: how do different
candidate broken-symmetry states compare at the same energy, particle
number, and slow observables? Maximizing the coarse-grained entropy
separately for each candidate provides a practical basis for this comparison.
It identifies entropically favored possibilities for further investigation,
while whether they actually form and persist remains a dynamical question.

For an isolated many-body system after a pulse, let $\hat H$ and $\hat N$
denote the Hamiltonian and particle-number operator, respectively, and let
$\hat\rho$ denote the many-body density operator describing its coarse-grained
state. The corresponding energy
$E=\mathrm{Tr}(\hat\rho\hat H)$ and particle number
$N=\mathrm{Tr}(\hat\rho\hat N)$ are constants of motion. Suppose, in
addition, that observables represented by operators $\hat Q_a$ relax on much
longer timescales. In the intermediate window
$\tau_{\mathrm{fast}}\ll t\ll\tau_{Q_a}$, their expectation values
$Q_a=\mathrm{Tr}(\hat\rho\hat Q_a)$ may therefore be treated as fixed. The
least biased coarse-grained state compatible with this retained information
is obtained by maximizing the von Neumann entropy
$-\mathrm{Tr}(\hat\rho\ln\hat\rho)$, where we set $k_{\mathrm B}=1$.
Introducing Lagrange multipliers for the constraints amounts to extremizing
\begin{equation}
\begin{split}
\mathcal{S}[\hat\rho]={}&-\mathrm{Tr}(\hat\rho\ln\hat\rho)
-\eta\bigl(\mathrm{Tr}\hat\rho-1\bigr)
-\beta\bigl[\mathrm{Tr}(\hat\rho\hat H)-E\bigr]\\
&-\alpha\bigl[\mathrm{Tr}(\hat\rho\hat N)-N\bigr]
-\sum_a\lambda_a
\bigl[\mathrm{Tr}(\hat\rho\hat Q_a)-Q_a\bigr].
\end{split}
\end{equation}
Here $\mathrm{Tr}$ denotes the trace over the many-body Hilbert space,
$\eta$ enforces normalization of $\hat\rho$, and $\beta$, $\alpha$, and
$\lambda_a$ enforce the prescribed values of $E$, $N$, and $Q_a$,
respectively.

If no information beyond $E$ and $N$ is retained, the last term is absent,
and stationarity gives the thermal ensemble
$\hat\rho_{\mathrm{th}}=\mathcal Z^{-1}
\exp[-\beta(\hat H-\mu\hat N)]$, where
$\mu=-\alpha/\beta$ is the chemical potential and $\mathcal Z$ normalizes
the density operator. In the thermodynamic limit, this ensemble is equivalent
to a microcanonical description at the corresponding energy and particle
number. Thermalization thus describes the loss of all preparation information
beyond these conserved quantities. If the $Q_a$ remain constrained, the same
variation instead gives
$\hat\rho_{\mathrm{rel}}=\mathcal Z^{-1}
\exp[-\beta(\hat H-\mu\hat N)-\sum_a\lambda_a\hat Q_a]$~\cite{SM}.
The retained values $Q_a$ distinguish quasisteady states with the same $E$
and $N$ and can therefore alter the order selected before complete
thermalization.

For each candidate order $\Phi$, we define
$S_\star(\Phi;E,N,{Q_a})$ as the largest entropy compatible with the
imposed constraints and select the macrostate that maximizes $S_\star$ over
$\Phi$. All candidates are compared at the same physical energy and retained
observables, although their Lagrange multipliers need not coincide. This
constrained entropy comparison is the variational principle used below.

\begin{figure}[tbp]
\centering
\paperfigure{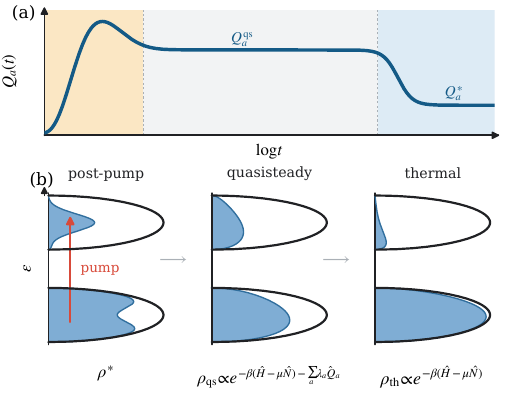}{0.80in}
\caption{Schematic illustration of the separation of relaxation times underlying nonequilibrium state selection. (a) Following excitation, a slow observable \(Q_a(t)\) rapidly approaches a long-lived quasisteady plateau \(Q_a^{\rm qs}\) before relaxing toward its thermal value \(Q_a^\ast\). (b) Schematic density of states and occupations immediately after pumping, after rapid intraband redistribution, and at thermal equilibrium. Blue shading denotes occupied spectral weight. The quasisteady ensemble maximizes the entropy at fixed energy, particle number, and \(Q_a\); subsequent relaxation of \(Q_a\) recovers the ordinary thermal ensemble.}
\label{fig:schematic}
\end{figure}

A direct optimization over interacting many-body states is generally
intractable because the Hilbert-space dimension grows exponentially with
system size. We make the calculation practical by restricting the density
matrix to number-conserving Gaussian states. Within this manifold, Wick
factorization produces a Hartree--Fock energy functional, and the
quasiparticle occupations and broken-symmetry fields are determined
self-consistently at fixed $E$, $N$, and $Q_a$~\cite{SM}. We call this implementation
variational nonequilibrium Hartree--Fock (VNHF). The Gaussian restriction is
an approximation used to evaluate the variational principle, which can in
principle be implemented with more accurate many-body descriptions.

The essential step beyond the relevant-ensemble construction is to vary the broken-symmetry macrostate itself at fixed retained observables. This construction builds on maximum-entropy and
relevant-ensemble formalisms, generalized Gibbs ensembles for exact
constraints, and prethermal descriptions based on approximate conservation
\cite{Jaynes1957,Mori1965,Zubarev1974,Rigol2007,Mori2018}.

The variational principle predicts statistical selection within the prescribed
class of states, rather than the dynamics by which the entropy-selected state is
reached. The relevant $\hat Q_a$, their lifetimes, and the kinetic
accessibility of the candidate orders, including nucleation and domain
growth, must be established experimentally or through a suitable dynamical
calculation.

We illustrate the framework with a few examples, beginning with a photoexcited Mott insulator. Doublon--holon pairs created across a large gap can be long lived because recombination requires transferring the gap energy to many low-energy excitations \cite{Strohmaier2010,EcksteinWerner2011,LenarcicPrelovsek2013}. Their population can therefore remain fixed after rapid intraband relaxation and provide a quasisteady constraint independent of energy, as also suggested by steady-state and generalized-ensemble descriptions of photodoped states \cite{LiEckstein2021,Murakami2022GGE}. Photoinduced metallization, nonthermal magnetism, and distinct charge, spin, lattice, and orbital relaxation further demonstrate why a single effective temperature may be insufficient \cite{Okamoto2011,Dean2016,Werner2012,Tsuji2013,Balzer2015,BakshiMajumdar2024,BakshiBoseDuttaMajumdar2024,RayWerner2024}.

For a tractable realization, we consider the half-filled Hubbard model
\cite{Hubbard1963},
$\hat H=-t_{\mathrm{hop}}\sum_{\langle ij\rangle\sigma}
(\hat c_{i\sigma}^\dagger\hat c_{j\sigma}+\mathrm{H.c.})
+U\sum_i\hat n_{i\uparrow}\hat n_{i\downarrow}$,
and evaluate its energy and entropy within magnetic Hartree--Fock theory.
This approximation implements the statistical principle; its ordered
branches are not intended as a complete description of a correlated Mott
spectrum. We represent the slow photocarrier sector by the upper-branch
population $n_+=N_+/N_s$, where $N_s$ is the number of sites. Maximizing the
fermionic entropy at fixed total energy and branch populations gives
\begin{equation}
f_{\bm k\nu}=
\begin{cases}
\displaystyle
\frac{1}{1+\exp[\beta(\varepsilon_{\bm k+}-\mu_+)]},
& \nu=+,\\[6pt]
\displaystyle
\frac{1}{1+\exp[\beta(\varepsilon_{\bm k-}-\mu_-)]},
& \nu=-.
\end{cases}
\label{eq:occupations}
\end{equation}
The derivation is given in the Supplemental Material~\cite{SM}. One total-energy
constraint gives a common temperature $T=\beta^{-1}$, while inhibited
particle transfer permits distinct chemical potentials $\mu_\pm$. This
assumes that energy is redistributed between the branches before their
populations relax; otherwise the separate branch energies must also be
retained. We set $k_{\mathrm B}=\hbar=1$.

The distribution and magnetic order are solved together using the
physical Hartree--Fock energy.
The band population is defined by projectors of the instantaneous ordered
Hamiltonian, and is an effective slow coordinate rather than an exact
microscopic doublon operator. Its conservation must be checked dynamically,
especially when the bands cease to be separated.
Throughout the displayed maps we restrict the intraband distributions to
$\beta\geq0$; gray regions have no admissible solution within that
restriction and the chosen order manifold.

\begin{figure}[tbp]
\centering
\paperfigure{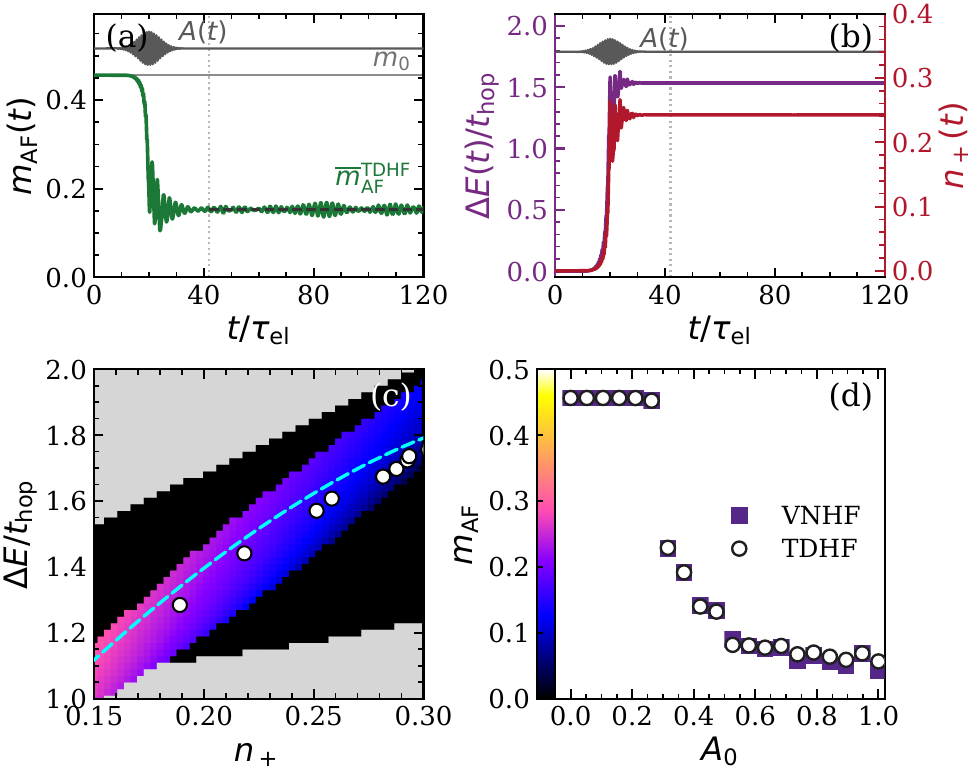}{9cm}
\caption{Square-lattice Hubbard model at $U/W=1.125$, calculated on a
$100\times100$ $\bm{k}$ grid.
(a) Pump-induced evolution of the Néel moment.
(b) Absorbed energy per site and instantaneous upper-band population
$n_+$; the gray pulse provides a time reference. The pump has a Gaussian envelope width
$\Omega=4t_{\mathrm{hop}}^{-1}$ and central frequency
$\omega_p=0.9W$.
(c) Constrained moment in the $(n_+,\Delta E)$ plane. The dashed curve
denotes the equilibrium trajectory, and the open circles mark the
states generated by the pump.  Gray regions contain no
admissible VNHF solution.
(d) Long-time TDHF moments compared with VNHF predictions at the same
energy and upper-band population.}
\label{fig:sqr_hubb}
\end{figure}

On the square lattice, VNHF predicts the long-time N\'eel moment using
only the absorbed energy and upper-branch population, in close agreement
with time-dependent Hartree--Fock (TDHF).
The moment
$m=|\langle(\hat n_{i\uparrow}-\hat n_{i\downarrow})/2\rangle|$
opens branches
$\varepsilon_{\bm k\pm}=\pm\sqrt{\epsilon_{\bm k}^{\,2}+(Um)^2}$,
where $\epsilon_{\bm k}=-2t_{\mathrm{hop}}(\cos k_x+\cos k_y)$ and the
common Hartree shift is omitted. At each $(n_+,\Delta E)$, with
$\Delta E=(E_{\mathrm{HF}}-E_{\mathrm{HF}}^{(0)})/N_s$, the constraints
determine $T$, $\mu_\pm$, and $m$~\cite{SM}; particle--hole symmetry gives
$\mu_-=-\mu_+$. Figure~\ref{fig:sqr_hubb} uses $U/W=1.125$ and
$W=8t_{\mathrm{hop}}$. The equilibrium curve, obtained on releasing the
population constraint, has $\mu_+=\mu_-$. Away from it, equal absorbed
energies can support different moments because excited carriers weaken
the imbalance between the two magnetic branches.

\begin{figure}[tbp]
\centering
\paperfigure{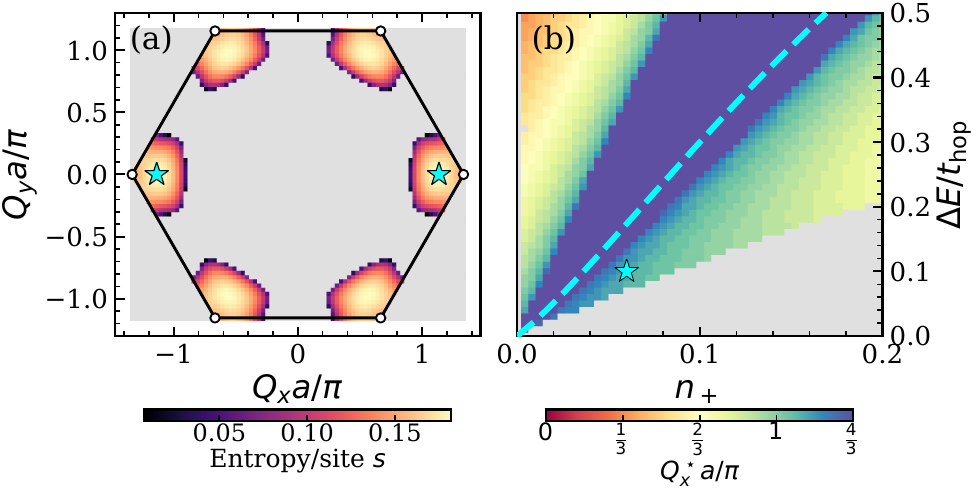}{1.30in}
\caption{Triangular-lattice Hubbard model at $U=6.5t_{\mathrm{hop}}$, calculated on a
$120\times120$ $\bm{k}$ grid.
(a) Constrained entropy versus spiral wave vector for a representative
excitation $(n_+=0.06,\Delta E=0.1 t_{\text{hop}})$, with maxima marked by stars.
(b) Entropy-selected $Q_x$ along $\bm Q=(Q_x,0)$ in the
$(n_+,\Delta E)$ plane. The $Q_xa/\pi=4/3$ plateau denotes
$120^\circ$ order; departures identify incommensurate spirals.
The dashed curve denotes equilibrium. Gray regions lack an admissible
solution within the stated search.}
\label{fig:tri_hubb}
\end{figure}

We benchmark this prediction against homogeneous TDHF, driven by a
Gaussian vector-potential pulse through the Peierls substitution.
The one-particle density matrix obeys
$i\partial_t\gamma=[h_{\mathrm{HF}}(t),\gamma]$.
The pulse suppresses the moment, while the post-pulse energy and
instantaneous upper-band population approach nearly stationary values
[Figs.~\ref{fig:sqr_hubb}(a,b)]. Their post-pulse averages, extracted for
each pump amplitude, supply the static calculation without fitting the
moment. Varying the pump amplitude traces a path in $(n_+,\Delta E)$
space distinct from the equilibrium curve [Fig.~\ref{fig:sqr_hubb}(c)].
Along this nonequilibrium path, the constrained prediction closely
follows the long-time TDHF moment, including its disappearance
[Fig.~\ref{fig:sqr_hubb}(d)]. This agreement concerns the order parameter:
collisionless TDHF preserves the eigenvalues of $\gamma$ and does not
establish the thermal occupations in Eq.~\eqref{eq:occupations}. It tests
whether the retained macroscopic quantities capture the dominant control
of magnetic melting, not equality of the two density matrices.

On the triangular lattice, frustration creates competing magnetic
structures \cite{WhiteChernyshev2007,Wietek2021, trilatt}, allowing the same
constraints to favor a different ordering wave vector. We choose
$U=6.5t_{\mathrm{hop}}$, where the Hartree--Fock ground state has
$120^\circ$ order, but an incommensurate spiral can maximize the entropy
after photoexcitation. To compare these states, we consider coplanar spirals
$\bm m_i=m[\cos(\bm Q\cdot\bm R_i)\hat{\bm x}
+\sin(\bm Q\cdot\bm R_i)\hat{\bm y}]$.
For each $\bm Q$, the two branches follow from a $2\times2$ Hamiltonian
with diagonal entries $\epsilon_{\bm k\pm\bm Q/2}$ and off-diagonal
entry $-Um$. We solve the population, energy, and moment constraints,
then compare the entropies. Here the chemical potentials must be
determined independently because particle--hole symmetry is absent.

The representative entropy landscape in Fig.~\ref{fig:tri_hubb}(a)
exhibits symmetry-related maxima. Tracking a representative along
$\bm Q=(Q_x,0)$ yields Fig.~\ref{fig:tri_hubb}(b): a commensurate region
at $Q_xa/\pi=4/3$, corresponding to $120^\circ$ order, gives way to
incommensurate spirals. Excitation reduces the branch occupation
imbalance and the self-consistent moment, weakening the exchange
splitting. Near competing spiral states, this changes how each wave
vector accommodates the retained carriers and energy. The entropy
maximum can therefore move away from the equilibrium $\bm K$ point
before magnetism disappears. This resembles access to competing orders
at smaller interaction strength, but is not a literal replacement of
$U$: the nonthermal populations remain essential.

This ordering tendency is consistent with real-space TDHF calculations,
which found excitation-induced spiral correlations and a long-lived
upper-band population~\cite{BakshiMondalMajumdar2024}. That work used a
nonequilibrium energy landscape at prescribed excitation. Here the
occupations follow from the entropy principle and candidate wave vectors
are compared at the same retained energy as well as population.
Nucleation, domain growth, and relaxation times still require dynamics.

The same construction also applies to open systems, where the energy
need not be conserved: a stationary energy maintained by driving and
dissipation can enter the entropy maximization on the same formal footing
as a retained $Q_a$. We consider the continuously driven double-exchange model
$\hat H_{\mathrm{DE}}=-t_{hop}\sum_{\langle ij\rangle s}
(\hat c_{is}^\dagger\hat c_{js}+\mathrm{H.c.})
-J_{\mathrm H}\sum_i\bm S_i\cdot
\hat c_i^\dagger\bm\sigma\hat c_i$,
with classical moments $|\bm S_i|=S$ and exchange splitting
$2J=2J_{\mathrm H}S$. Driven simulations of this model exhibit a conversion
from metallic FM order to N\'eel correlations
\cite{OnoIshihara2017,OnoIshihara2018,OnoIshihara2020}. In the regime of
interest, slow relaxation between the two Hund branches prevents their
populations from fully equilibrating. Previous simulations primarily access
the limit $n_+\simeq0$; below we also examine a finite retained upper-branch
population.

\begin{figure}[tbp]
\centering
\paperfigure{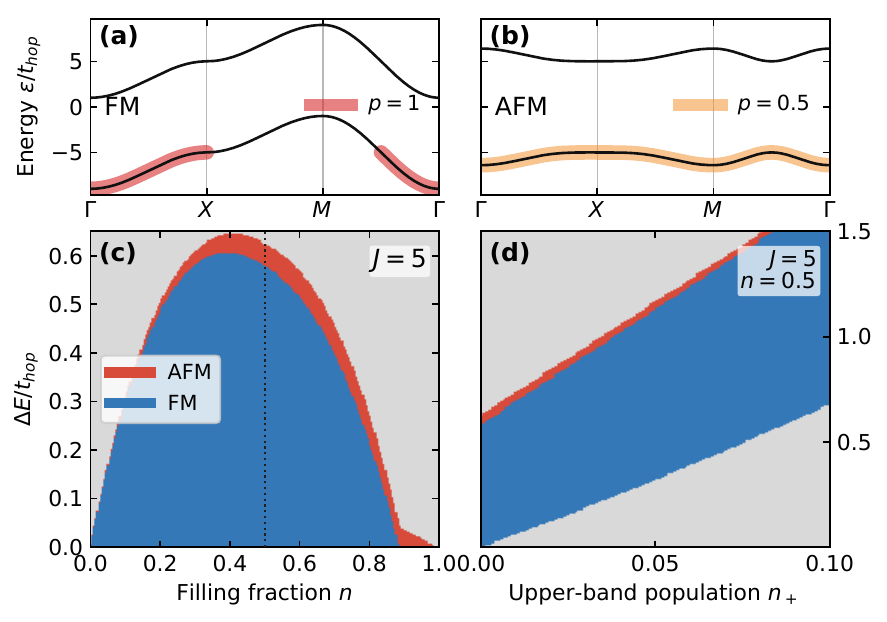}{1.45in}
\caption{Conditional order selection in the square-lattice double-exchange
model. (a,b) Electronic bands of the FM and Néel-AFM textures at filling
$n=1/2$. Thick colored segments indicate representative occupations: the
zero-temperature Fermi-sea distribution of the FM lower band in (a), and the
uniform occupation $f_{\bm k-}=1/2$ of the AFM lower band in (b); the upper
band is empty in both cases. (c) Texture with the larger constrained
electronic entropy versus filling $n$ and excess energy
$\Delta E=\bar e-e_{\mathrm{gs}}(n)$ at fixed $n_+=0$. The dotted line marks
$n=1/2$. (d) Corresponding comparison at fixed $n=1/2$ as the upper-band
population $n_+$ is varied, with $n_-=1/2-n_+$ and $\Delta E$ measured from
the $n=1/2$, $n_+=0$ ground state. All panels use
$J/t_{\mathrm{hop}}=5$, where $J=J_{\mathrm H}S$. Blue and red denote
FM- and AFM-favored regions, respectively, while gray lies outside the common
domain of admissible noninverted intraband distributions. In a driven
realization, the steady-state energy and branch populations entropy-select the point
sampled from these maps.}
\label{fig:f2af}
\end{figure}

Here $\dot e=P_{\mathrm{in}}-P_{\mathrm{loss}}$, so a nonequilibrium steady
state (NESS) has a stationary time-averaged energy $\bar e$ fixed by
$\overline{P}_{\mathrm{in}}=\overline{P}_{\mathrm{loss}}$.
At each prescribed $(\bar e,n,n_+)$, with $n_-=n-n_+$, we maximize the
electronic entropy within the FM and N\'eel textures and compare the
resulting values. Figure~\ref{fig:f2af}(c) considers $n_+=0$ over a range
of fillings, whereas Fig.~\ref{fig:f2af}(d) fixes $n=1/2$ and varies $n_+$.
The dynamics determines which point $(\bar e,n,n_+)$ is realized, while
the constrained entropy comparison determines which texture is favored
there. Because the electronic Hamiltonian is already quadratic for a
prescribed spin texture, no Hartree--Fock decoupling or self-consistency
is required.

The AFM-favored regions provide an entropic explanation for the driven
FM-to-AFM conversion. At $n=1/2$ and $n_+=0$, the uniform AFM occupation
$f_{\bm k-}=1/2$ maximizes the lower-band entropy, giving $s=\ln 2$, whereas
an FM state at the same energy generally requires a nonuniform occupation
and therefore has lower entropy. This advantage persists over the finite
energy window shown in Fig.~\ref{fig:f2af}(c). For nonzero $n_+$, the
maximum possible intraband entropy is $h(n_-)+h(n_+)$, where
$h(x)=-x\ln x-(1-x)\ln(1-x)$, but the distinct FM and AFM dispersions
reach a given energy with different occupation profiles. Consequently,
increasing $n_+$ shifts the conditional FM--AFM boundary, as shown in
Fig.~\ref{fig:f2af}(d). The realized energy, populations, and
classical-spin dynamics remain determined by the drive and dissipation
and can depend strongly on the damping \cite{OnoIshihara2020}.

Here, we have developed a constrained maximum-entropy framework for identifying nonequilibrium quasisteady order in systems with well-separated relaxation times. After rapidly relaxing degrees of freedom lose detailed memory of the preparation, the remaining state is characterized by particle number, energy in a closed system, and a small set of slowly relaxing observables. Restricting the density matrix to the Gaussian manifold gives a practical variational nonequilibrium Hartree--Fock construction in which the occupations and broken-symmetry fields are determined self-consistently. The underlying principle is nevertheless more general and is not tied to the Hartree--Fock approximation.

In the square-lattice Hubbard model, the absorbed energy and the long-lived upper-band population together account for the pump-induced suppression of antiferromagnetism and closely reproduce the long-time TDHF moment. On the triangular lattice, the same constraints reshape the entropy landscape and shift the preferred ordering wave vector from the equilibrium $120^\circ$ state to an incommensurate spiral. The double-exchange example provides the complementary open-system setting: a drive-generated population confined to the lower Hund band can favor an antiferromagnetic electronic texture, although the NESS energy and its dynamical accessibility are ultimately determined by the balance between driving and dissipation. Thus, slowly relaxing populations can control not only whether an ordered state survives, but also which order is entropy-selected.

The framework predicts statistical compatibility and relative preference, rather than the kinetics of state formation. It does not identify the relevant slow observables, determine their lifetimes, or describe nucleation, domain growth, coarsening, and dynamical barriers \cite{chern_cdw1,chern_orbital_ordering,chern_phase_separation,BakshiMajumdar2024}; these must be established by experiment or an appropriate dynamical calculation. The present Hartree--Fock implementation neglects collective fluctuations, strong correlations, and spatial coexistence.

The central advance is that, once the slow observables and a prescribed class of candidate states are specified, the search for nonequilibrium order becomes a static constrained optimization. One can scan magnetic structures, ordering wave vectors, anomalous channels, or spatially modulated states and compare their constrained entropies without performing a separate long-time simulation for every preparation protocol. This makes the method an efficient screening tool for locating promising hidden or transient orders and identifying where more expensive dynamical calculations and experiments should be focused. 
More broadly, this construction provides a nonequilibrium analogue of equilibrium variational state selection: slow observables replace temperature alone as the macroscopic information that determines which ordered state is statistically favored.

\vspace{1cm}

\begin{acknowledgments}
\textit{Acknowledgements:} This work was supported by the US Department of Energy Basic Energy Sciences under Contract No. DE-SC0020330. The authors acknowledge Research Computing at the University of Virginia for providing computational resources and technical support.
\end{acknowledgments}

\bibliography{refs}

\clearpage
\onecolumngrid
\appendix

\begin{center}
{\large\bfseries Supplemental Material for}\\[0.5em]
{\large\bfseries
``Nonequilibrium Broken-Symmetry States from Retained Observables''}\\[0.8em]
{\normalsize
Sankha Subhra Bakshi and Gia-Wei Chern}\\[0.4em]
{\small
Department of Physics, University of Virginia,\\
Charlottesville, Virginia 22904, USA}
\end{center}

\vspace{1em}

\twocolumngrid

\setcounter{equation}{0}
\renewcommand{\theequation}{A\arabic{equation}}
\renewcommand{\theHequation}{appendix.\arabic{equation}}

\section{Constrained ensemble and Gaussian Hartree--Fock reduction}

Consider a statistical state constrained by the physical energy
$E=\mathrm{Tr}(\hat\rho\hat H)$, particle number
$N=\mathrm{Tr}(\hat\rho\hat N)$, and expectation values
$Q_a=\mathrm{Tr}(\hat\rho\hat Q_a)$ of additional slowly relaxing
observables. The least biased state compatible with this information
maximizes the von Neumann entropy
$S_{\mathrm{vN}}[\hat\rho]=-\mathrm{Tr}(\hat\rho\ln\hat\rho)$.
Introducing Lagrange multipliers gives the functional
\begin{equation}
\begin{split}
\mathcal S[\hat\rho]={}&S_{\mathrm{vN}}[\hat\rho]
-\eta\bigl(\mathrm{Tr}\hat\rho-1\bigr)
-\beta\bigl[\mathrm{Tr}(\hat\rho\hat H)-E\bigr]\\
&-\alpha\bigl[\mathrm{Tr}(\hat\rho\hat N)-N\bigr]
-\sum_a\lambda_a
\bigl[\mathrm{Tr}(\hat\rho\hat Q_a)-Q_a\bigr].
\end{split}
\label{eq:appendix_entropy_functional}
\end{equation}
Using
$\delta S_{\mathrm{vN}}
=-\mathrm{Tr}[\delta\hat\rho(\ln\hat\rho+1)]$,
its first variation is
\begin{equation}
\begin{split}
\delta\mathcal S=-\mathrm{Tr}\,\delta\hat\rho
\biggl[&\ln\hat\rho+1+\eta+\beta\hat H
+\alpha\hat N
+\sum_a\lambda_a\hat Q_a\biggr].
\end{split}
\label{eq:appendix_entropy_variation}
\end{equation}
Because $\delta\hat\rho$ is arbitrary, stationarity requires
\begin{equation}
\ln\hat\rho
=-(1+\eta)-\beta\hat H-\alpha\hat N
-\sum_a\lambda_a\hat Q_a .
\end{equation}
Exponentiating this relation and imposing
$\mathrm{Tr}\hat\rho=1$ gives
\begin{equation}
\hat\rho_{\mathrm{rel}}
=\frac{1}{\mathcal Z}
\exp\left[
-\beta(\hat H-\mu\hat N)
-\sum_a\lambda_a\hat Q_a
\right],
\qquad
\mu=-\frac{\alpha}{\beta},
\label{eq:appendix_relevant}
\end{equation}
where $\mathcal Z$ normalizes the density operator. If no additional
$Q_a$ are retained, Eq.~\eqref{eq:appendix_relevant} reduces to the
thermal ensemble fixed by $E$ and $N$. Retaining the $Q_a$ instead
produces a quasisteady ensemble that preserves information not contained
in the conserved energy and particle number. 

For an interacting Hamiltonian,
Eq.~\eqref{eq:appendix_relevant} is generally non-Gaussian. We therefore
approximate the constrained maximization by restricting $\hat\rho$ to
the manifold of number-conserving fermionic Gaussian states,
\begin{equation}
\hat\rho_{\mathrm G}
=\frac{1}{\mathcal Z_{\mathrm G}}
\exp\left(-\sum_{ij}K_{ij}
\hat c_i^\dagger\hat c_j\right).
\label{eq:appendix_gaussian_rho}
\end{equation}
Such a state is completely determined by its one-particle density matrix
\begin{equation}
\gamma_{ij}
=\mathrm{Tr}\bigl(\hat\rho_{\mathrm G}
\hat c_j^\dagger\hat c_i\bigr),
\qquad
\boldsymbol{\gamma}
=\bigl(1+e^{\boldsymbol K}\bigr)^{-1}.
\label{eq:appendix_gamma}
\end{equation}
Its entropy is
\begin{equation}
S_{\mathrm G}[\boldsymbol{\gamma}]
=-\mathrm{tr}_1\left[
\boldsymbol{\gamma}\ln\boldsymbol{\gamma}
+(1-\boldsymbol{\gamma})
\ln(1-\boldsymbol{\gamma})\right],
\label{eq:appendix_gaussian_entropy}
\end{equation}
where $\mathrm{tr}_1$ denotes a trace over the one-particle Hilbert
space. If $f_\ell$ are the eigenvalues of $\boldsymbol{\gamma}$, this
becomes
\begin{equation}
S_{\mathrm G}
=\sum_\ell h(f_\ell),
\qquad
h(f)=-f\ln f-(1-f)\ln(1-f).
\label{eq:appendix_binary_entropy}
\end{equation}

To see how the Gaussian restriction generates Hartree--Fock theory,
consider a general two-body Hamiltonian
\begin{equation}
\hat H
=\sum_{ij}t_{ij}\hat c_i^\dagger\hat c_j
+\frac{1}{2}\sum_{ijkl}V_{ij;kl}
\hat c_i^\dagger\hat c_j^\dagger
\hat c_l\hat c_k .
\label{eq:appendix_two_body_hamiltonian}
\end{equation}
Wick's theorem gives
\begin{equation}
\left\langle
\hat c_i^\dagger\hat c_j^\dagger
\hat c_l\hat c_k
\right\rangle_{\mathrm G}
=\gamma_{ki}\gamma_{lj}
-\gamma_{li}\gamma_{kj},
\label{eq:appendix_wick}
\end{equation}
and hence
\begin{equation}
\begin{split}
E_{\mathrm G}[\boldsymbol{\gamma}]
={}&\sum_{ij}t_{ij}\gamma_{ji}\\
&+\frac{1}{2}\sum_{ijkl}V_{ij;kl}
\bigl(\gamma_{ki}\gamma_{lj}
-\gamma_{li}\gamma_{kj}\bigr).
\end{split}
\label{eq:appendix_gaussian_energy}
\end{equation}
The two contractions in the second line are respectively the direct
and exchange contributions. The functional derivative of this
Gaussian energy defines the Hartree--Fock single-particle Hamiltonian,
\begin{equation}
(h_{\mathrm{HF}})_{ij}
\equiv
\frac{\partial E_{\mathrm G}}
{\partial\gamma_{ji}},
\qquad
\delta E_{\mathrm G}
=\mathrm{tr}_1
\bigl(\boldsymbol h_{\mathrm{HF}}
\delta\boldsymbol{\gamma}\bigr).
\label{eq:appendix_hf_derivative}
\end{equation}
Equivalently, the interacting Hamiltonian is replaced on the Gaussian
manifold by the self-consistent quadratic Hamiltonian
\begin{equation}
\hat H_{\mathrm{HF}}[\boldsymbol{\gamma}]
=E_{\mathrm c}[\boldsymbol{\gamma}]
+\sum_{ij}(h_{\mathrm{HF}})_{ij}
\hat c_i^\dagger\hat c_j,
\label{eq:appendix_hf_hamiltonian}
\end{equation}
with
\begin{equation}
E_{\mathrm c}
=E_{\mathrm G}
-\mathrm{tr}_1
\bigl(\boldsymbol h_{\mathrm{HF}}
\boldsymbol{\gamma}\bigr).
\label{eq:appendix_double_counting}
\end{equation}
Thus $E_{\mathrm c}$ removes the interaction energy counted twice in
the Hartree--Fock eigenvalue sum.

For the slow observables considered below,
$\hat Q_a=\sum_{ij}(q_a)_{ij}\hat c_i^\dagger\hat c_j$, so that
$Q_a=\mathrm{tr}_1(\boldsymbol q_a\boldsymbol{\gamma})$. The constrained
Gaussian functional is therefore
\begin{equation}
\begin{split}
\mathcal S_{\mathrm G}
={}&S_{\mathrm G}
-\beta\bigl(E_{\mathrm G}-E\bigr)
-\alpha\bigl(\mathrm{tr}_1\boldsymbol{\gamma}-N\bigr)\\
&-\sum_a\lambda_a
\bigl[\mathrm{tr}_1
(\boldsymbol q_a\boldsymbol{\gamma})-Q_a\bigr].
\end{split}
\label{eq:appendix_gaussian_functional}
\end{equation}
The entropy variation is
\begin{equation}
\delta S_{\mathrm G}
=\mathrm{tr}_1\left[
\ln\left(\frac{1-\boldsymbol{\gamma}}
{\boldsymbol{\gamma}}\right)
\delta\boldsymbol{\gamma}\right].
\label{eq:appendix_gaussian_entropy_variation}
\end{equation}
Consequently, stationarity with respect to
$\boldsymbol{\gamma}$ gives
\begin{equation}
\ln\left(\frac{1-\boldsymbol{\gamma}}
{\boldsymbol{\gamma}}\right)
=\beta\boldsymbol h_{\mathrm{HF}}[\boldsymbol{\gamma}]
+\alpha\boldsymbol 1
+\sum_a\lambda_a\boldsymbol q_a ,
\label{eq:appendix_gaussian_stationarity}
\end{equation}
or
\begin{equation}
\boldsymbol{\gamma}
=\left\{
1+\exp\left[
\beta\bigl(\boldsymbol h_{\mathrm{HF}}
-\mu\boldsymbol 1\bigr)
+\sum_a\lambda_a\boldsymbol q_a
\right]\right\}^{-1}.
\label{eq:appendix_constrained_hf}
\end{equation}
Equations~\eqref{eq:appendix_hf_derivative} and
\eqref{eq:appendix_constrained_hf} must be solved together. This is the
constrained finite-temperature Hartree--Fock problem: the Gaussian
density matrix determines $\boldsymbol h_{\mathrm{HF}}$, while
$\boldsymbol h_{\mathrm{HF}}$ determines the Gaussian density matrix.

For the two-branch applications, let
$\boldsymbol h_{\mathrm{HF}}\lvert\ell\nu\rangle
=\varepsilon_{\ell\nu}\lvert\ell\nu\rangle$, with $\nu=\pm$, and fix
the two branch populations
$N_\nu=\sum_\ell f_{\ell\nu}$. The two population constraints produce
distinct multipliers $\alpha_\pm$, or equivalently distinct chemical
potentials $\mu_\pm=-\alpha_\pm/\beta$. In the Hartree--Fock eigenbasis,
Eq.~\eqref{eq:appendix_constrained_hf} becomes
\begin{equation}
\begin{aligned}
f_{\ell+}
&=\frac{1}{
1+\exp[\beta(\varepsilon_{\ell+}-\mu_+)]},\\[3pt]
f_{\ell-}
&=\frac{1}{
1+\exp[\beta(\varepsilon_{\ell-}-\mu_-)]}.
\end{aligned}
\label{eq:appendix_fd}
\end{equation}
The single energy constraint produces a common multiplier \(\beta\) for the two branches. In the isolated applications, \(T=\beta^{-1}\), whereas in the NESS application \(\beta\) is simply the multiplier conjugate to the prescribed stationary energy.
The separately fixed branch populations
produce the two chemical potentials $\mu_+$ and $\mu_-$. The remaining
constraints are
\begin{equation}
N_\nu=\sum_\ell f_{\ell\nu},
\qquad
E=E_{\mathrm c}
+\sum_{\ell,\nu}
\varepsilon_{\ell\nu}f_{\ell\nu}.
\label{eq:appendix_constraints}
\end{equation}
For a prescribed spin texture in the double-exchange model, the
single-particle Hamiltonian is independent of
$\boldsymbol{\gamma}$ and $E_{\mathrm c}=0$. In the Hubbard
applications, both $\boldsymbol h_{\mathrm{HF}}$ and
$E_{\mathrm c}$ depend on the self-consistent density matrix.

For a fixed spectrum, the strict concavity of $h(f)$ makes
Eq.~\eqref{eq:appendix_fd} the unique entropy-maximizing occupation
function for feasible interior constraints. The full Hartree--Fock
problem can nevertheless possess several self-consistent solutions
because $E_{\mathrm G}[\boldsymbol{\gamma}]$ is nonlinear. We therefore
solve for all candidate ordered states satisfying the same physical
$E$, $N$, and $Q_a$, evaluate their Gaussian entropies, and select the
state with the largest $S_{\mathrm G}$.

\section{Square-lattice Néel state}

For the half-filled square-lattice Hubbard model, define
\begin{equation}
R_{\bm k}
=\sqrt{\epsilon_{\bm k}^{\,2}+(Um_{\mathrm{AF}})^2},
\qquad
\varepsilon_{\bm k\pm}=\pm R_{\bm k},
\end{equation}
where
$\epsilon_{\bm k}=-2t_{\mathrm{hop}}(\cos k_x+\cos k_y)$.
Using the reduced Brillouin zone (RBZ), the population, energy, entropy,
and self-consistency equations are
\begin{align}
n_\nu
&=\frac{2}{N_s}
\sum_{\bm k\in\mathrm{RBZ}}f_{\bm k\nu},
\qquad n_-+n_+=1,
\label{eq:appendix_square_number}\\
e_{\mathrm{HF}}
&=\frac{2}{N_s}
\sum_{\substack{\bm k\in\mathrm{RBZ}\,\nu=\pm}}
\varepsilon_{\bm k\nu}f_{\bm k\nu}
+Um_{\mathrm{AF}}^2+\frac{U}{4},
\label{eq:appendix_square_energy}\\
s_{\mathrm G}
&=\frac{2}{N_s}
\sum_{\substack{\bm k\in\mathrm{RBZ}\,\nu=\pm}}
h(f_{\bm k\nu}),
\label{eq:appendix_square_entropy}\\
1
&=\frac{U}{N_s}
\sum_{\bm k\in\mathrm{RBZ}}
\frac{f_{\bm k-}-f_{\bm k+}}{R_{\bm k}},
\qquad m_{\mathrm{AF}}>0.
\label{eq:appendix_square_gap}
\end{align}
The factor of two accounts for spin degeneracy. The constant $U/4$
restores the uniform Hartree contribution associated with the original
Hubbard Hamiltonian. It cancels from the excess energy
$\Delta E=e_{\mathrm{HF}}-e_{\mathrm{HF}}^{(0)}$, whereas the
double-counting contribution $Um_{\mathrm{AF}}^2$ must be retained.
The paramagnetic solution $m_{\mathrm{AF}}=0$ is evaluated separately
and included in the entropy comparison. At half filling,
particle--hole symmetry gives
$\mu_-=-\mu_+$ and
$f_{\bm k-}=1-f_{\bm k+}$.

\section{Triangular-lattice spiral state}

For the triangular lattice,
\begin{equation}
\epsilon_{\bm k}
=-2t_{\mathrm{hop}} 
\left[
\cos k_x
+2\cos \left(\frac{k_x}{2}\right)
\cos \left(\frac{\sqrt{3}k_y}{2}\right)
\right].
\end{equation}
For a coplanar spiral with ordering wave vector $\bm Q$, define
\begin{align}
\bar\epsilon_{\bm k\bm Q}
&=\frac{
\epsilon_{\bm k+\bm Q/2}
+\epsilon_{\bm k-\bm Q/2}}{2},\nonumber\\
\xi_{\bm k\bm Q}
&=\frac{
\epsilon_{\bm k+\bm Q/2}
-\epsilon_{\bm k-\bm Q/2}}{2},\nonumber\\
R_{\bm k\bm Q}
&=\sqrt{\xi_{\bm k\bm Q}^{\,2}+(Um)^2}.
\end{align}
The two spiral branches are
\begin{equation}
\varepsilon_{\bm k\pm}
=\bar\epsilon_{\bm k\bm Q}\pm R_{\bm k\bm Q}.
\end{equation}
Each branch contains $N_s$ states in the full Brillouin zone. The
closure equations are therefore

\begin{align}
n_\nu
&=\frac{1}{N_s}\sum_{\bm k}f_{\bm k\nu},
\qquad n_-+n_+=1,
\label{eq:appendix_triangular_number}\\
e_{\mathrm{HF}}
&=\frac{1}{N_s}\sum_{\bm k,\nu}
\varepsilon_{\bm k\nu}f_{\bm k\nu}
+Um^2+\frac{U}{4},
\label{eq:appendix_triangular_energy}\\
s_{\mathrm G}
&=\frac{1}{N_s}\sum_{\bm k,\nu}
h(f_{\bm k\nu}),
\label{eq:appendix_triangular_entropy}\\
1
&=\frac{U}{2N_s}\sum_{\bm k}
\frac{f_{\bm k-}-f_{\bm k+}}{R_{\bm k\bm Q}},
\qquad m>0.
\label{eq:appendix_triangular_gap}
\end{align}

Because the triangular lattice is not particle--hole symmetric,
$\mu_-$ and $\mu_+$ are determined independently. For every trial
$\bm Q$, Eqs.~\eqref{eq:appendix_fd} and
\eqref{eq:appendix_triangular_number}--\eqref{eq:appendix_triangular_gap}
are solved at the prescribed $(e_{\mathrm{HF}},n_+)$. The entropies of
all converged solutions are then compared. The ordering wave vector
reported in the main text maximizes $s_{\mathrm G}$ within the
single-$\bm Q$ coplanar-spiral family and the stated symmetry-line
search; the remaining symmetry-related wave vectors follow from the
sixfold lattice symmetry.

\section{Double-exchange textures}

For prescribed FM and N\'eel-AFM classical-spin textures, the electronic
Hamiltonian is quadratic and requires no self-consistency. With
$J=J_{\mathrm H}S$, the two branches are
\begin{align}
\varepsilon_{\bm k\pm}^{\mathrm{FM}}
&=\epsilon_{\bm k}\pm J,\nonumber\\
\varepsilon_{\bm k\pm}^{\mathrm{AF}}
&=\pm\sqrt{\epsilon_{\bm k}^{2}+J^2},
\label{eq:appendix_de_bands}
\end{align}
where each branch contains $N_s$ single-particle states.

For a texture $X\in\{\mathrm{FM},\mathrm{AF}\}$, the maximum-entropy
occupations at fixed branch populations and energy are
\begin{equation}
f_{\ell\nu}^{X}
=\frac{1}{
\exp[\beta_X(\varepsilon_{\ell\nu}^{X}-\mu_\nu^{X})]+1},
\qquad \nu=\pm ,
\label{eq:appendix_de_occupations}
\end{equation}
with closure equations
\begin{align}
n_\nu
&=\frac{1}{N_s}\sum_{\ell=1}^{N_s}f_{\ell\nu}^{X},
\qquad n_-+n_+=n,
\label{eq:appendix_de_number}\\
e
&=\frac{1}{N_s}\sum_{\ell,\nu}
\varepsilon_{\ell\nu}^{X}f_{\ell\nu}^{X},
\label{eq:appendix_de_energy}\\
s_X
&=\frac{1}{N_s}\sum_{\ell,\nu}
h(f_{\ell\nu}^{X}).
\label{eq:appendix_de_entropy}
\end{align}
For each prescribed $(e,n,n_+)$, these equations are solved separately
for the FM and AFM textures, and their entropies are compared. The case
$n_+=0$ is obtained by setting $f_{\ell+}^{X}=0$.

For the noninverted distributions used in the phase maps
$(\beta_X\geq0)$, let the levels within each branch be ordered increasingly
and define $N_\nu=n_\nu N_s$. The accessible energy interval is
\begin{align}
e_0^X(n_-,n_+)
&=\frac{1}{N_s}\sum_{\nu=\pm}
\sum_{\ell=1}^{N_\nu}\varepsilon_{\ell\nu}^{X},
\nonumber\\
e_{\mathrm{unif}}^X(n_-,n_+)
&=\frac{1}{N_s}\sum_{\ell,\nu}
n_\nu\varepsilon_{\ell\nu}^{X},
\nonumber\\
e_0^X(n_-,n_+)
&\leq e\leq e_{\mathrm{unif}}^X(n_-,n_+).
\label{eq:appendix_de_interval}
\end{align}
The upper limit corresponds to uniform occupations
$f_{\ell\nu}^{X}=n_\nu$ and entropy
$s_{\max}=h(n_-)+h(n_+)$. Entropies are compared only within the common
accessible interval of the two textures; points outside this interval are
shown in gray. Figure~\ref{fig:f2af}(c) uses $n_+=0$, while
Fig.~\ref{fig:f2af}(d) fixes $n=1/2$ and varies $n_+$.

\end{document}